\documentclass{article}
\usepackage{graphicx}

\newcommand{\tdots}{\mathinner{\ldotp\ldotp}}

\begin{document}

\title{Multiple Sequence Alignment Based on Set Covers}

\author{Alexandre H. L. Porto\\
Valmir~C.~Barbosa\thanks{Corresponding author ({\tt valmir@cos.ufrj.br}).}\\
\\
Universidade Federal do Rio de Janeiro\\
Programa de Engenharia de Sistemas e Computa\c c\~ao, COPPE\\
Caixa Postal 68511\\
21941-972 Rio de Janeiro - RJ, Brazil}

\date{December 10, 2004}

\maketitle

\begin{abstract}
We introduce a new heuristic for the multiple alignment of a set of sequences. The heuristic is based on a set 
cover of the residue alphabet of the sequences, and also on the determination of a significant set of blocks comprising 
subsequences of the sequences to be aligned. These blocks are obtained with the aid of a new data structure, called a 
suffix-set tree, which is constructed from the input sequences with the guidance of the residue-alphabet set cover and generalizes
the well-known suffix tree of the sequence set. 
We provide performance results on selected BAliBASE amino-acid sequences and compare them with those yielded by some 
prominent approaches.\\

\end{abstract}

\section{Introduction}

The multiple alignment of biological sequences is one of the most important problems in computational molecular biology, 
with applications in many different important domains, such as the analysis of protein structure and function, the 
detection of conserved patterns and domain organization of a protein family, evolutionary studies based on phylogenetic 
analysis, and database searching for new members of a protein family. For recent reviews highlighting the importance of multiple 
alignments in molecular biology, we refer the reader to \cite{g97, g99, ltptp01, ejt04}.

The problem of multiple sequence alignment can be stated in the following manner. Let $s_1,\ldots,s_k$, with $k\geq 2$, be 
sequences of lengths $n_1,\ldots,n_k$, respectively, over a residue alphabet $R$. An alignment $\mathcal{A}$ of these 
sequences is a $k\times l$ matrix such that $\mathcal{A}[i,j]$, for $1\leq i\leq k$ and $1\leq j\leq l$, is either a 
character in $R$ or the special character that we call a gap character. In $\mathcal{A}$, fixing $i$ and varying $j$ from $1$ 
through $l$ must reproduce $s_i$ exactly if gap characters are skipped. Fixing $j$, in turn, yields $k$ characters that are 
said to be aligned in $\mathcal{A}$, of which at least one must be in $R$. Note, then, that 
$\max\{n_1,\ldots,n_k\}\leq l\leq n_1+\cdots+n_k$.

The goal of the multiple sequence alignment problem \cite{s75, wsb76} is to determine the most biologically significant 
alignment of $s_1,\ldots,s_k$. Finding this alignment requires an objective function to associate a score with each 
possible alignment, and in this case the multiple sequence alignment problem is to find an alignment, 
known as the optimal alignment, that maximizes the objective function. There exist many different objective functions 
that can be used \cite{g99}, but none of them guarantees that the corresponding optimal alignment is the most 
biologically significant alignment of the sequences \cite{n02}.

It follows from the definition of an alignment that the number of different alignments of a given set of sequences is 
exponentially large \cite{dms98, s98}; in fact, the multiple sequence alignment problem is known to be NP-hard 
\cite{wj94, bv01, j01, m03}. Feasible approaches to solve the problem are then all of a heuristic nature, as can be seen in 
\cite{g97, sm97, dekm98, g99, p00, bb01, nrd02, n02, ejt04}.

In this paper we describe a new heuristic that is based on set covers of the residue alphabet $R$. Such a set cover is a 
collection of subsets of $R$ whose union yields $R$ precisely. The idea behind the use of a set cover is that each subset can 
be made to contain all the residues from $R$ that possess certain structural or physicochemical properties in common
\cite{t86, t99, vvp01}. The most familiar scenario for the use of set covers is the case of $R$ as a set of amino acids, so 
henceforth we concentrate mainly on multiple protein alignments, even though it makes perfect sense for $R$ to be a set of
nucleotides as well. Set covers of an amino-acid alphabet have been studied extensively, as for example in 
\cite{s66, jz81, t86, ss90, lb93, m95, k96b, nfjwn96, s96, wb96, i97, ctrv02, lfww03}.

Set covers lie at the heart
of the new heuristic.  In essence, what they do is to allow the introduction of a new data structure, called a 
suffix-set tree, that generalizes the well-known suffix tree of a set of sequences and can be used in the determination of 
subsequence blocks that ultimately give rise to the desired alignment. In general terms, this is the same approach as 
some others in the literature \cite{sm86, wj90, mbrzh94, zhm96, pfr99, zj01}, but our use of set covers as its basis 
provides a fundamentally more direct link between relevant properties shared by groups of residues and the resulting 
alignment.

The following is how the remainder of the paper is organized. In Section~\ref{sec-newtree} we introduce our new data 
structure and in Section~\ref{sec-newheur} describe our new approach to multiple sequence alignment. Then we proceed in 
Section~\ref{sec-testheur} to an experimental study of the new method as compared to some of its most prominent competitors, 
and finalize with conclusions in Section~\ref{sec-concrem}.

\section{Suffix-set trees} \label{sec-newtree}

In this section we describe our new data structure. It is a generalization of the well-known suffix tree of a set of
sequences \cite{w73, g97}, which is one of the most important data structures in the field of pattern recognition. Such a suffix 
tree has $O(n_1+\cdots+n_k)$ nodes and can be constructed in $O(n_1+\cdots+n_k)$ time by several algorithms \cite{w73, m76, u95, gs97, g97}.

Suffix trees can be applied to many problems, but their principal application in computational molecular biology is to 
assist the algorithms that try to obtain blocks comprising biologically significant subsequences of a set of sequences, 
known as the motifs of that set \cite{d81}. These motifs, in the case of proteins, encode structural or functional 
similarities; in the case of nucleic acids, they encode mainly the promoter regions. There exist several algorithms for 
extracting motifs from a set of sequences, based on a multitude of different algorithmic and mathematical methodologies. 
We refer the reader to \cite{bjeg98, rfpgp00, sw03} in the general case, and to \cite{vms99} in the case of nucleic acids.

\subsection{Definition and properties}

Let ${\mathcal{C}}=\{C_1,\ldots,C_p\}$ be a set cover of $R$, and let 
$\Sigma_{\mathcal{C}}=\{\alpha_1,\ldots,\alpha_p,\alpha_{\$}\}$ be a new alphabet having a character $\alpha_i$ for each 
$C_i\in{\mathcal{C}}$ and a further character $\alpha_{\$}$ possessing a function similar to the terminator used in suffix 
trees. Like the suffix tree, our new data structure is also a rooted tree; it has edges labeled by sequences of characters from 
$\Sigma_{\mathcal{C}}$ and nodes labeled by indices into some of $s_1,\ldots,s_k$ to mark suffix beginnings.
We call it a suffix-set tree and it has the following properties:
\begin{itemize}
\item The first character of the label on an edge connecting a node to one of its children is a different character of 
$\Sigma_{\mathcal{C}}$ for each child.
\item Each nonempty suffix of every one of the $k$ sequences is associated with at least one leaf of the tree; conversely, each leaf of the tree is 
associated with at least one nonempty suffix of some sequence (if more that one, then all suffixes associated with the leaf have 
the same length).\footnote{These properties may alternatively be extended to accommodate the empty suffix as well.
We refrain from doing so for simplicity's sake only.}
Thus, each leaf is labeled by a set like $\left\{(i_1,j_1),\ldots,(i_q,j_q)\right\}$ for some $q\ge 1$,
where $(i_a,j_a)$, for $1\leq a\leq q$, $1\leq i_a\leq k$, and $1\leq j_a\leq n_{i_a}$,
indicates that the suffix $s_{i_a}[j_a\tdots n_{i_a}]$ of $s_{i_a}$ is associated with the leaf.
\item Let $v$ be a node of the tree. The label of $v$ is the set $\left\{(i_1,j_1),\ldots,(i_q,j_q)\right\}$ that represents the $q$
suffixes collectively associated with the leaves of the subtree rooted at $v$. If $\alpha_{c_1}\cdots\alpha_{c_r}$ is the 
concatenation of the labels on the path from the root of the tree to $v$, excluding if necessary the terminal character 
$\alpha_{\$}$, then $\alpha_{c_1}\cdots\alpha_{c_r}$ is a common representation of all prefixes of length $r$ of the 
suffixes associated with the leaves of the subtree rooted at $v$. If $s_{i_a}[j_a\tdots n_{i_a}]$ is one of these 
suffixes, then for $1\le b\le r$ we have $s_{i_a}[j_a+b-1]\in C_{c_b}$ (that is, the $b$th character of the suffix is a member of 
$C_{c_b}$).
\end{itemize}

We discuss an example shortly, but let us first examine a simple procedure to construct a suffix-set tree. Before that, though,
the reader should note that, when $\mathcal{C}$ is a partition of $R$ into singletons, the suffix-set tree becomes the familiar 
suffix tree of $s_1,\ldots,s_k$. In order to see this, it suffices to identify for each character in each sequence the member $C_i$ of
$\mathcal{C}$ to which it belongs, and then substitute $\alpha_i$ for that character.\footnote{We include edge labels in our definition of
the suffix-set tree only on account of this aspect of it as a generalizer of the suffix tree. Edge labels are nowhere strictly necessary in this
paper's application of the suffix-set tree, but there may exist other applications that make use of them, just like the case with
suffix trees.}

\subsection{A simple construction algorithm}

We now describe an algorithm to construct the suffix-set tree of a set of sequences. The algorithm generates a tree $T$ 
with $n_T$ nodes having the properties described above. For each suffix of each sequence, the algorithm first processes its
first character, then its second character, and so on. We describe it as the following three steps:
\begin{enumerate}
\item Create a root $r$ for $T$ and label it with a set comprising every possible pair $(i,j)$, i.e., for $1\leq i\leq k$ and 
$1\leq j\leq n_i$. Let $v=r$ and $c=1$.  \label{algtree-step-1}
\item Let $\left\{(i_1,j_1),\ldots,(i_q,j_q)\right\}$ be the label of $v$. Let $D_C$, for every $C\in{\mathcal{C}}$, 
and $D_{\$}$ be (initially empty) sets of node-label elements. For each $(i_a,j_a)$ in the label of $v$, check whether 
$j_a+c-1\leq n_{i_a}$. In the affirmative case (i.e., the suffix $s_{i_a}[j_a\tdots n_{i_a}]$ has at least $c$ 
characters), add $(i_a,j_a)$ to each $D_C$ such that $s_{i_a}[j_a+c-1]\in C$; in the negative case, add 
$(i_a,j_a)$ to $D_{\$}$. With ${\mathcal{C}}'=\{C\in{\mathcal{C}}\mid D_C\neq\emptyset\}$ initially, while 
there exist $C,C'\in{\mathcal{C}}'$ such that $D_C\subseteq D_{C'}$, do 
${\mathcal{C}}'={\mathcal{C}}'\setminus\{C\}$.  \label{algtree-step-2} 
\item If ${\mathcal{C}}'=\emptyset$, $D_{\$}\neq\emptyset$, and $v\neq r$, then append $\alpha_{\$}$ to the label of 
the edge connecting $v$ to its parent. If ${\mathcal{C}}'=\{C_i\}$, $D_{\$}=\emptyset$, and $v\neq r$, then 
append $\alpha_i$ to the label of the edge connecting $v$ to its parent and go to Step~\ref{algtree-step-2} with $c+1$ in 
place of $c$. Otherwise, create a new leaf $\ell$ and the edge $e$ between $v$ and $\ell$ if $D_{\$}\neq\emptyset$, then label 
$\ell$ with $D_{\$}$ and $e$ with the character $\alpha_{\$}$. Also, for each $C_i\in{\mathcal{C}}'$, create a new node 
$v_{C_i}$ and the edge $e_{C_i}$ between $v$ and $v_{C_i}$, then label $v_{C_i}$ with $D_{C_i}$, $e_{C_i}$ with $\alpha_i$, and go to 
Step~\ref{algtree-step-2} with $v=v_{C_i}$ and $c+1$ in place of $c$. \label{algtree-step-3}
\end{enumerate}

Notice that, in Step~\ref{algtree-step-2}, we eliminate from ${\mathcal{C}}'$ every $C$ for which 
$C'\in{\mathcal{C}}'$ exists such that $D_C\subseteq D_{C'}$. This implies that sets in 
$\mathcal{C}$ that are subsets of other sets also in $\mathcal{C}$ become useless. Whenever this is for some reason 
undesirable, the pruning of ${\mathcal{C}}'$ may be replaced by a simple elimination of duplicates (yielding, 
necessarily, a larger suffix-set tree).

\begin{figure}
\begin{center}
\includegraphics[scale=0.92]{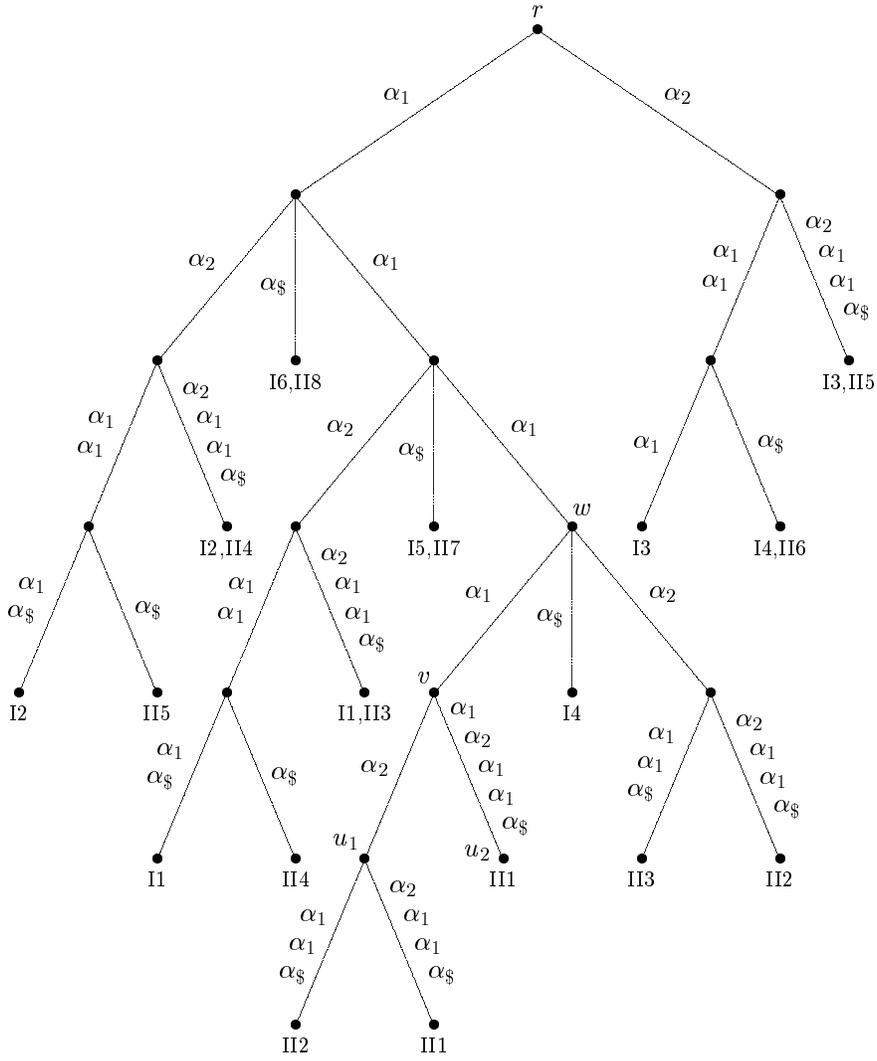}
\caption{An example suffix-set tree, node labels shown only for the leaves. Each node label is expressed more compactly than
in the text; for example, ``I2,II4'' stands for the label $\left\{(1,2),(2,4)\right\}$.}
\label{fig-1}
\end{center}
\end{figure}

We show in Figure~\ref{fig-1} the suffix-set tree obtained by this latter alternative for $R=\{A,C,G,T\}$,
$C_1=\{A,G,T\}$, $C_2=\{T,C\}$, $s_1=AGCTAG$, and $s_2=GGGATCGA$. Had Steps~\ref{algtree-step-1}--\ref{algtree-step-3} been used as 
presented, the result would have been essentially the same, except for the inexistence of nodes $v$ and $u_2$ (and all 
adjoining edges) and the addition of an edge between $w$ and $u_1$.

If we recall that $p$ is the number of sets in $\mathcal{C}$, then it is relatively simple to see that
Steps~\ref{algtree-step-1}--\ref{algtree-step-3} run in $O\left(n_T\left(p(n_1+\cdots+n_k)+p^2|R|^2\right)\right)$ time and require 
$O\left(n_T(n_1+\cdots+n_k)\right)$ space. In these figures, the time complexity is essentially dominated by
Step~\ref{algtree-step-2}, the space complexity by the need to store node labels. What is worrisome in both cases is the linear 
dependence on the number of nodes of the suffix-set tree, $n_T$, since it is conceivable that $n_T$ may itself depend exponentially 
on $\max\{n_1,\ldots,n_k\}$ for some sequence sets and set covers in the worst case. 

\begin{figure}
\begin{center}
\begin{tabular}{c}
\includegraphics[scale=0.55]{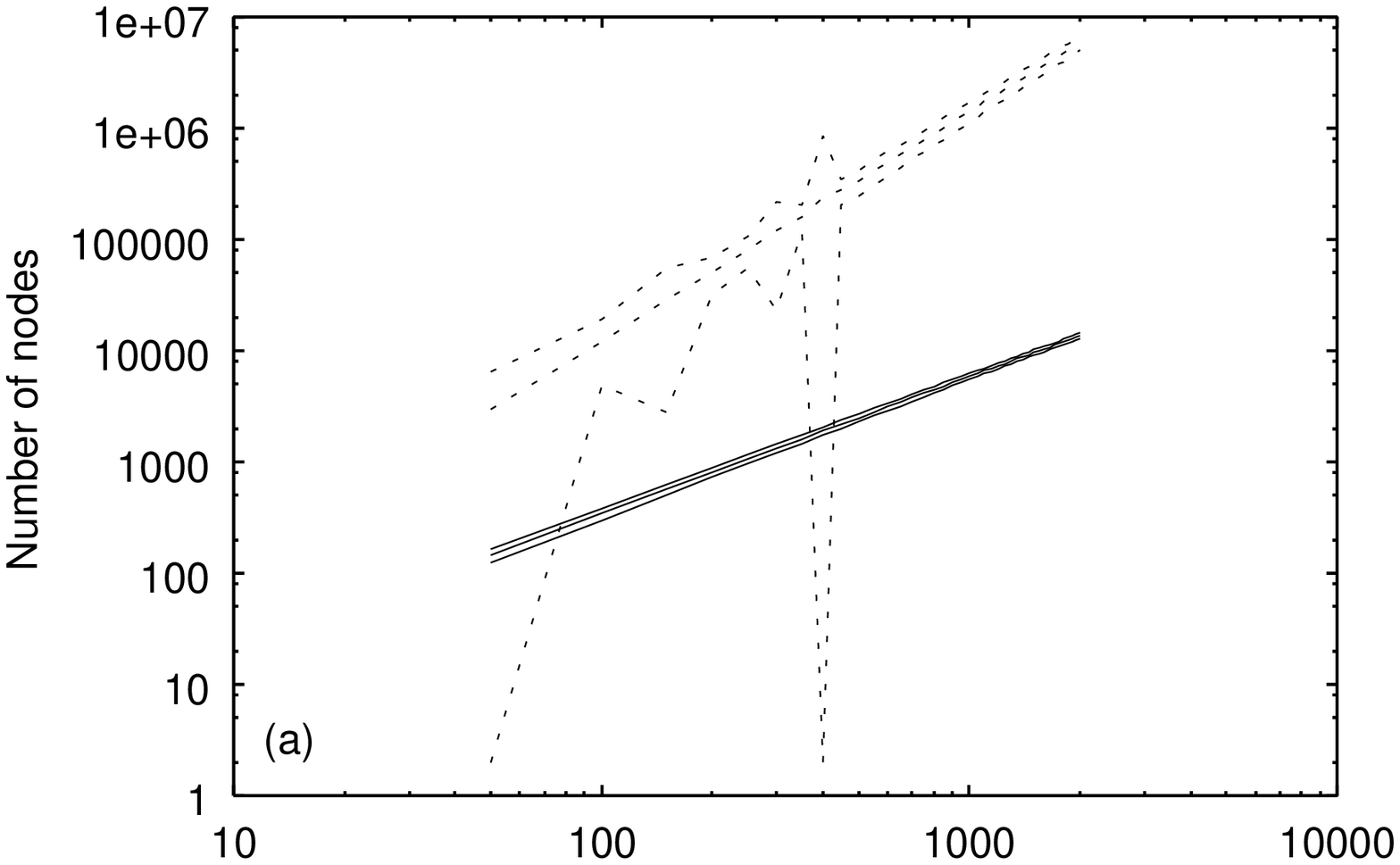} \\
\\
\includegraphics[scale=0.55]{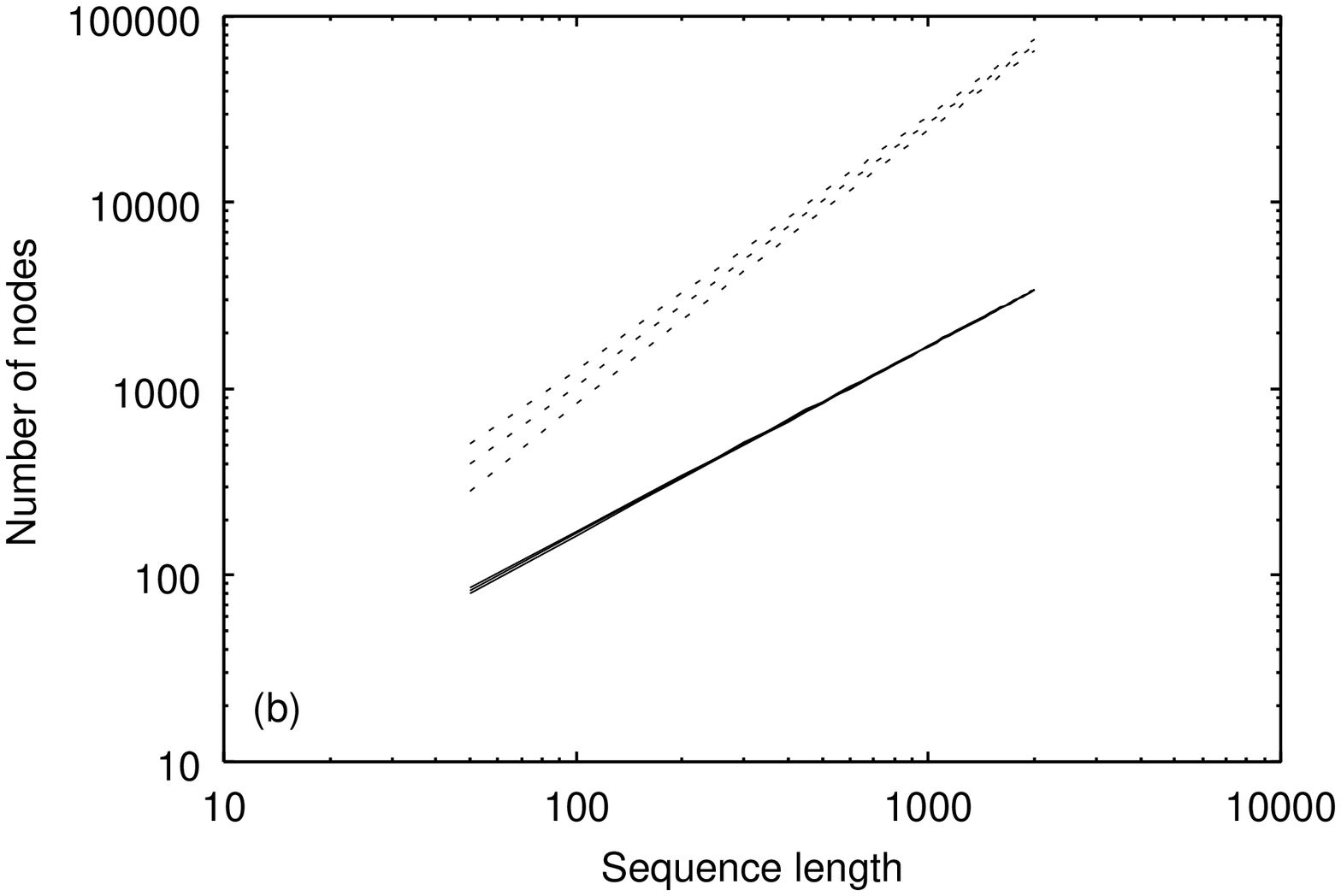} \\
\end{tabular}
\caption{Expected behavior of $n_T$ for single-sequence suffix-set trees under the $\mathcal{I}$ (a) or $\mathcal{S}$ (b) set cover.}
\label{fig-2}
\end{center}
\end{figure}

Let us then probe a little more deeply into the nature of $n_T$. One first important observation is that, notwithstanding 
such a worst-case possibility, we have found by means of experiments that in many biologically relevant cases the expected 
value of $n_T$ grows polynomially, not exponentially, with $n_1+\cdots+n_k$. The results of these experiments are shown in the plots of 
Figure~\ref{fig-2}, in which part (a) refers to the set cover from \cite{i97}, here denoted by $\mathcal{I}$, and part 
(b) refers to the set cover from \cite{ss90}, here denoted by $\mathcal{S}$. In both cases, $R$ is the set of amino 
acids; the set covers are given in Table~\ref{tab-1}. In the figure, solid plots refer to
Steps~\ref{algtree-step-1}--\ref{algtree-step-3} as presented and give the average value (by itself and plus or minus the 
standard deviation) of $n_T$ over single-sequence suffix-set trees for $10^3$ randomly generated sequences.\footnote{Is is curious to 
notice, in Figure~\ref{fig-2}(b), that the three solid plots practically overlap one another. This happens because the 
set pruning performed at the end of Step~\ref{algtree-step-2} turns the set cover $\mathcal{S}$ into a partition of $R$. 
Not every subset of $R$ in this partition is a singleton, so the suffix-set tree that is constructed is not the suffix
tree of the sequence on $R$. Nevertheless, the suffix-set tree can be regarded as the suffix tree of the single sequence
on $\Sigma_{\mathcal{S}}$ that corresponds by trivial substitution to the sequence on $R$.}
Dashed plots, correspondingly, refer to the 
alternative version of the tree-constructing algorithm. In all cases, it is clear from the plots' linearity with respect to
the two logarithmic scales that, on average, $n_T$ depends only polynomially on the sequence's length. Not only this, but 
it also seems true that deviations above the average occur only in modest amounts.

\begin{table}
\begin{center}
\caption{Amino-acid set covers.}
\begin{tabular}{lll}
\hline
Cover set & $\mathcal{I}$ & $\mathcal{S}$ \\
\hline
$C_1$ & $\{M,I,L,V\}$ & $\{P\}$ \\
$C_2$ & $\{M,I,L,V,A,P\}$ & $\{A,G\}$ \\
$C_3$ & $\{M,I,L,V,F,W\}$ & $\{D,E\}$ \\
$C_4$ & $\{M,I,L,V,A,P,F,W\}$ & $\{N,Q\}$ \\
$C_5$ & $\{D,E,H,R,K\}$ & $\{S,T\}$ \\
$C_6$ & $\{S,T,Q,N\}$ & $\{F,W,Y\}$ \\ 
$C_7$ & $\{S,T,Q,N,D,E\}$ & $\{H,K,R\}$ \\
$C_8$ & $\{Q,N,D,E,H,R,K\}$ & $\{I,L,V\}$ \\
$C_9$ & $\{S,T,Q,N,D,E,H,R,K\}$ & $\{C,F,I,L,M,V,W,Y\}$ \\
$C_{10}$ & $\{Q,N\}$ & $\{D,E,H,K,N,Q,R,S,T\}$ \\
$C_{11}$ & $\{D,E,Q,N\}$ & \\
$C_{12}$ & $\{H,R,K\}$ & \\
$C_{13}$ & $\{R,K\}$ &  \\
$C_{14}$ & $\{F,W,Y\}$ & \\
$C_{15}$ & $\{G,N\}$ & \\
$C_{16}$ & $\{A,C,G,S\}$ & \\
$C_{17}$ & $\{S,T\}$ & \\
$C_{18}$ & $\{D,E\}$ & \\
\hline
\end{tabular}
\label{tab-1}
\end{center}
\end{table}

The second important observation is related to our preferred use of the tree-constructing algorithm in Section~\ref{sec-newheur}. 
In the best-performing options of the strategy to be described in that section (cf.\ Section~\ref{sec-testheur} as well),
we do not construct the suffix-set tree to completion, but 
rather only the portion of the tree that is needed to represent all suffix prefixes of length at most $M$ (a fixed 
parameter). Achieving this requires a simple modification to Step~\ref{algtree-step-2}: we add the pair $(i_a,j_a)$ to 
the $D_C$ sets if $j_a+c-1\leq n_{i_a}$ and in addition $c\leq M$. Clearly, we now have that $n_T$ is $O(p^{M})$, therefore 
polynomial in $p$ given the fixed parameter $M$.

\section{Alignments from set covers} \label{sec-newheur}

For $2\leq k'\leq k$, let $t_1,\ldots,t_{k'}$ be subsequences of
$s_1,\ldots,s_k$ such that each $t_a$ is a subsequence of a different $s_{i_a}$, with $1\leq a\leq k'$ and $1\leq i_a\leq k$. We call 
$\{t_1,\ldots,t_{k'}\}$ a block. If  $\mathcal{A}'$ is an alignment of $t_1,\ldots,t_{k'}$ 
having $l'$ columns, then the score of  $\mathcal{A}'$, denoted by $S(\mathcal{A}')$, is
a function, to be introduced shortly, of
\begin{equation}
T(\mathcal{A}')=\sum_{c=1}^{l'}\sum_{a=1}^{k'-1}\sum_{b=a+1}^{k'}Q\left(p^c_a,p^c_b,\mathcal{A}'[a,c],\mathcal{A}'[b,c]\right),
\label{eq-score1}
\end{equation}
where $p_a^c$ is the position in $s_{i_a}$ of the rightmost character of $t_a$ whose column in 
$\mathcal{A}'$ is no greater than $c$ ($p_a^c=0$, if none exists), and similarly for $p_b^c$.
In (\ref{eq-score1}), $Q\left(p^c_a,p^c_b,\mathcal{A}'[a,c],\mathcal{A}'[b,c]\right)$
is the contribution to $T(\mathcal{A}')$ of aligning the two characters $\mathcal{A}'[a,c]$ and 
$\mathcal{A}'[b,c]$ given $p_a^c$ and $p_b^c$. If both $\mathcal{A}'[a,c]$ and 
$\mathcal{A}'[b,c]$ are gap characters, then we let $Q\left(p^c_a,p^c_b,\mathcal{A}'[a,c],\mathcal{A}'[b,c]\right)=0$.
Otherwise, the value of $Q\left(p^c_a,p^c_b,\mathcal{A}'[a,c],\mathcal{A}'[b,c]\right)$ 
is determined through a sequence of two conceptual steps.

The first step is the determination of the combined number of optimal global and local pairwise alignments of $s_{i_a}$ 
and $s_{i_b}$ that go through the $p_a^c$,$p_b^c$ cell of the dynamic-programming matrix. We dwell very briefly on how this
number can be found, mentioning only that, even though it can be exponentially large \cite{dms98, s98}, a simple
extension of the procedure of \cite{nw70} in the global case or of \cite{sw81} in the local case suffices to compute it
efficiently. In what follows, we split this number into three others, and let
each of $U_1(p_a^c,p_b^c),\ldots,U_3(p_a^c,p_b^c)$ be either a linearly normalized version of the corresponding number
within the interval $[1,L]$ for $L$ a parameter,
if that number is strictly positive, or $0$, otherwise. We use $U_1(p_a^c,p_b^c)$ in reference to the case of
optimal alignments through the $p_a^c$,$p_b^c$ cell that align $s_{i_a}[p_a^c]$ to $s_{i_b}[p_b^c]$,
$U_2(p_a^c,p_b^c)$ for alignments of $s_{i_a}[p_a^c]$ to a gap character, and
$U_3(p_a^c,p_b^c)$ for alignments of $s_{i_b}[p_b^c]$ to a gap character.

The second step is the determination of $Q\left(p^c_a,p^c_b,\mathcal{A}'[a,c],\mathcal{A}'[b,c]\right)$ itself from the quantities
$U_1(p_a^c,p_b^c),\ldots,U_3(p_a^c,p_b^c)$. If $U_1(p_a^c,p_b^c)=\cdots=U_3(p_a^c,p_b^c)=0$ (no optimal alignments through the
$p_a^c$,$p_b^c$ cell), then we let
$Q\left(p^c_a,p^c_b,\mathcal{A}'[a,c],\mathcal{A}'[b,c]\right)=-L$. Otherwise,
we have the following cases to consider, where $z\in\{1,2,3\}$ selects among $U_1$, $U_2$, and $U_3$ depending, as explained above, on
$\mathcal{A}'[a,c]$ and $\mathcal{A}'[b,c]$:
\begin{itemize}
\item If $U_z(p_a^c,p_b^c)>0$, then we let $Q\left(p^c_a,p^c_b,\mathcal{A}'[a,c],\mathcal{A}'[b,c]\right)=U_z(p_a^c,p_b^c)$.
\item If $U_z(p_a^c,p_b^c)=0$, then we let 
\[
Q\left(p^c_a,p^c_b,\mathcal{A}'[a,c],\mathcal{A}'[b,c]\right)=-{\textstyle\min^+}\left\{U_1(p_a^c,p_b^c),U_2(p_a^c,p_b^c),U_3(p_a^c,p_b^c)\right\},
\]
where we use $\min^+$ to denote the minimum of the strictly positive arguments only.
\end{itemize}
What this second step is doing is to favor the alignment of $\mathcal{A}'[a,c]$ to $\mathcal{A}'[b,c]$ 
in proportion to its popularity in optimal pairwise alignments of $s_{i_a}$ and $s_{i_b}$, and similarly to penalize 
it---heavily when cell $p_a^c,p_b^c$ is part of no optimal pairwise alignment, less so if it is but not aligning 
$\mathcal{A}'[a,c]$ to $\mathcal{A}'[b,c]$.

Finally, the function that yields $S(\mathcal{A}')$ from $T(\mathcal{A}')$ is designed to
differentiate two alignments of different blocks for which $T$ might yield the same value.
We do so by subtracting off $T(\mathcal{A}')$ the fraction of $\vert T(\mathcal{A}')\vert$ obtained from the 
average of two numbers in $[0,1]$. The first number is $1-k'/k$ and seeks to privilege (decrease $T$ by a smaller value) the block with the greater 
number of subsequences. The second number is a function of the so-called identity score of an alignment, that is, the fraction of the 
number of aligned residue pairs that corresponds to identical residues. If we denote the identity score of 
$\mathcal{A}'$ by $I(\mathcal{A}')$, then the second number is $1-I(\mathcal{A}')$ and aims at privileging alignments whose identity scores
are comparatively greater. We then have
\begin{equation}
S(\mathcal{A}')=T(\mathcal{A}')-\left(2-\frac{k'}{k}-I(\mathcal{A}')\right)\frac{\vert T(\mathcal{A}')\vert}{2}.
\label{eq-score2}
\end{equation}

The remainder of this section is devoted to describing our heuristic to obtain a $k\times l$ alignment $\mathcal{A}$ 
of the sequences $s_1,\ldots,s_k$, given the set cover $\mathcal{C}$ of the residue alphabet $R$. The suffix-set tree $T$ of 
$s_1,\ldots,s_k$ plays a central role, and we assume it only represents suffix prefixes whose lengths do not exceed $M$, the parameter
introduced in Section~\ref{sec-newtree}.
We first describe how to obtain a suitable set $\mathcal{B}$ of blocks from $s_1,\ldots,s_k$, and then how to 
obtain $\mathcal{A}$ from $\mathcal{B}$.

\subsection{Blocks from set covers} \label{sec-newheur-blocks}

We start by creating a set $\mathcal{B}$ of blocks that is initialized to $\emptyset$ and is augmented by the inclusion of 
new blocks as they are generated. The size of $\mathcal{B}$ is at all times bounded by yet another parameter, denoted by $N$. 

Every node $v$ of $T$ that is not the root may contribute blocks to $\mathcal{B}$. If $n_v\leq k$ is the number of distinct 
sequences with suffixes associated with $v$ and $l_v\leq M$ is the common prefix length of all those suffixes, then each 
block contributed by $v$ is formed exclusively by some of the length-$l_v$ prefixes as its subsequences, totaling at least 
two and including at most one from each of the $n_v$ sequences. 

Let $\mathcal{A}_B$ denote an alignment of block $B$'s subsequences. Block formation for node $v$ proceeds as 
follows. First the $n_v$ sequences are sorted in nonincreasing order of their numbers of suffixes associated with $v$ and a 
new block is created for each prefix of the first-ranking sequence. An attempt is then made to add more prefixes to each such block 
$B$ by visiting the remaining $n_v-1$ sequences in order and selecting for each one the prefix, if any, that increases 
$S(\mathcal{A}_B)$ the most when $\mathcal{A}_B$ acquires another row by the addition of that prefix. It is worthy of 
mention that, as prefixes coalesce into the final form of $B$, $\mathcal{A}_B$ never contains any gap characters, in which case 
$Q$ is seen to revert to a simpler form. But notice that the functional form in
(\ref{eq-score2}) continues to be effective, not only because of the identity scores, but also because the expansion of $B$
involves comparing alignments that may differ in numbers of rows (a unit difference, to be specific).\footnote{In a similar 
vein, comparing candidate new rows based solely on their effect on $S(\mathcal{A}_B)$ may at times be insufficient. In our 
current implementation we use the simpler $T(\mathcal{A}_B)$ as well if needed to help break ties.} At the end, all blocks still having
one single sequence are discarded.

Once $v$'s contribution to $\mathcal{B}$ is available, it is sorted and then merged into $\mathcal{B}$ (we can think of
$\mathcal{B}$ as being internally organized as a sorted list). Both operations seek 
to retain inside $\mathcal{B}$ those blocks whose alignments' scores are greater. If needed, additional tie-breaking criteria 
are employed, including those that are already reflected in (\ref{eq-score2}): greater numbers of 
subsequences and identity scores are preferred.

Now say that two blocks $B$ and $B'$ are such that $B$ is contained in $B'$ if every subsequence of $B$ is itself 
a subsequence of the corresponding subsequence of $B'$. Once every non-root node of $T$ has been considered for contributing to 
$\mathcal{B}$, the resulting $\mathcal{B}$ is further pruned by the removal of every block that is contained in another of 
at least the same score. After this, $\mathcal{B}$ undergoes another fix, which is to extend its blocks by the addition of 
new subsequences so that at the end they all contain exactly one subsequence from each of $s_1,\ldots,s_k$.

The following is how we achieve this extension for block $B\in{\mathcal{B}}$. We consider the unrepresented sequences one by 
one in nonincreasing order of their lengths. Then we use a variation of a semi-global algorithm for aligning a sequence to 
a subsequence of another sequence \cite{g97, sm97} to align the current $\mathcal{A}_B$ to a subsequence of the unrepresented sequence 
under consideration. This variation is straightforward and employs the
$Q$ function in place of a substitution matrix and gap costs. As $B$ is thus 
extended, its alignment $\mathcal{A}_B$ changes as well, and may now acquire gap characters for the first time.

The final step in this setup process of $\mathcal{B}$ is to once again examine all its blocks and remove every block $B$ such 
that either $S(\mathcal{A}_B)<0$ or $\mathcal{A}_B$ is contained in $\mathcal{A}_{B'}$ for some other 
$B'\in{\mathcal{B}}$ for which $S(\mathcal{A}_B)\leq S(\mathcal{A}_{B'})$.\footnote{One technicality here is that, since all blocks have at
this point the same number $k$ of subsequences, (\ref{eq-score2}) no longer holds as given but rather as the simpler
form in which the $1-k'/k$ term is done away with and furthermore the division by $2$ is likewise not performed.} The containment of one alignment in 
another is a notion completely analogous to that of block containment introduced above.

\subsection{Alignments from blocks} \label{sec-newheur-alignment}

The $\mathcal{B}$ that we now have contains $k$-subsequence blocks exclusively, all having nonnegative-score alignments that 
are not contained in  one another (except when the container has a lower score). In this new phase of the heuristic we build 
a weighted acyclic directed graph $D$ from $\mathcal{B}$. Manipulating this graph appropriately eventually yields the 
desired alignment $\mathcal{A}$ of $s_1,\ldots,s_k$.

The node set of $D$ is ${\mathcal{B}}\cup\{s,t\}$, where $s$ and $t$ are two special nodes. In $D$, an edge exists directed 
from $s$ to each $B\in{\mathcal{B}}$, and also from each $B\in{\mathcal{B}}$ to $t$. No other edges are incident to $s$ or $t$,
which are then a source and a sink in $D$, respectively (i.e., edges only outgo from $s$ or income to $t$).
The additional edges of $D$ are deployed among the 
members of $\mathcal{B}$ in the following manner. For $B,B'\in{\mathcal{B}}$, an edge exists directed from $B$ to 
$B'$ if every subsequence of $B$ starts to the left of the corresponding subsequence in $B'$ in the 
appropriate sequence of $s_1,\ldots,s_k$. In addition, if $B$ and $B'$ overlap, then $\mathcal{A}_B$ and 
$\mathcal{A}_{B'}$ are also required to be identical in all the overlapping columns. Edges deployed in this manner 
lead, clearly, to an acyclic directed graph.

In $D$, both edges and nodes have weights. Edge weights depend on how the blocks intersect the sequences $s_1,\ldots,s_k$. 
Specifically, if an edge exists from $B$ to $B'$ and the two blocks are nonoverlapping, then its weight is $-x$, where $x$ is the 
standard deviation of the intervening sequence-segment lengths. Edges outgoing from $s$ or incoming to $t$ are weighted in 
the trivially analogous manner.

Weights for edges between overlapping blocks and node weights are computed similarly to each other (except for $s$ and $t$,
whose weights are equal to $0$). If $x$ is the number of 
residues in node $B$, then its weight is $x/\sqrt{k}$. In the case of an edge between the overlapping $B$ and $B'$, 
we let $x$ be the number of common residues and set the edge's weight to $-x/\sqrt{k}$. We remark, finally, that this 
weight-assignment methodology is very similar to the one in \cite{zj01}, the main difference being that we count residues 
instead of alignment sizes.

Having built $D$, we are then two further steps away from the final alignment $\mathcal{A}$. The first step is to find an 
$s$-to-$t$ directed path in $D$ whose weighted length is greatest. Since $D$ is acyclic, this can be achieved efficiently 
\cite{clrs01}. Every block $B$ appearing on this optimal path immediately contributes $\mathcal{A}_B$ as part of 
$\mathcal{A}$, but there still remain unaligned sequence segments.

The second step and final action of the heuristic is then to complete the missing positions of $\mathcal{A}$.
We describe what is done between nonoverlapping successive blocks, but clearly the same has to be applied to the left of the first
block on the optimal path and to the right of the last block.
Let $B$ and $B'$ be nonoverlapping blocks appearing in succession on the optimal path. Let $t_1,\ldots,t_k$ be the intervening 
subsequences of $s_1,\ldots,s_k$ that are still unaligned. We select the largest of $t_1,\ldots,t_k$ and use it to initialize 
a new alignment along with as many gap characters as needed for every one of $t_1,\ldots,t_k$ that is empty. We then visit 
each of the remaining subsequences in nonincreasing length order and align it to the current, partially built new alignment. 
The method used here is totally analogous to the one used in Section~\ref{sec-newheur-blocks} for providing every block with 
exactly $k$ subsequences, the only difference being that a global (as opposed to semi-global) procedure is used.

\section{Computational results} \label{sec-testheur}

We have conducted extensive experimentation in order to evaluate the performance of the heuristic of Section 
\ref{sec-newheur}. Our strategy has been to employ the BAliBASE suite \cite{tpp99a, bttp01} as the source of sequence sets, 
and to seek comparative results vis-\`a-vis some prominent approaches, namely CLUSTAL W \cite{thg94}, PRRN \cite{g96}, 
DIALIGN \cite{mfdw98, m99}, T-COFFEE \cite{nhh00}, and MAFFT \cite{kmkm02}. Some of these have already been the subject of 
comparative studies as a group \cite{tpp99b}, while some others constitute more recent proposals.

The BAliBASE suite comprises $167$ families of amino-acid sequences divided into eight reference sets, each of which especially 
constructed to emphasize some of the most common scenarios related to multiple sequence alignment. The suite contains a 
reference alignment for each of its families, in most cases along with motif annotations given the reference alignment. We 
have concentrated our experiments on the families for which such annotations are available, namely the first five reference 
sets \cite{bttp01}. 

The metrics we use to evaluate a certain alignment $\mathcal{A}$ are motif-constrained versions of those originally introduced along with the 
BAliBASE suite \cite{tpp99b}:\footnote{We do point out, however, that alternative methodologies have also been used 
\cite{kh01, ls02}.} the sum-of-pairs score (SPS), denoted by $SPS(\mathcal{A})$, and the column score (CS), denoted by 
$CS(\mathcal{A})$. If we let $p_{abc}$ be a $0$-$1$ variable such that $p_{abc}=1$ if and only if the residues 
$\mathcal{A}[a,c]$ and $\mathcal{A}[b,c]$ share the same column in the BAliBASE reference alignment for the same 
sequences, then the version of $SPS(\mathcal{A})$ we use is the average value of $p_{abc}$ over the residue pairs that are annotated as belonging to motifs 
in the reference alignment. Similarly, if $C_c$ is the $0$-$1$ variable having value $1$ if and only if all the residues in 
column $c$ of $\mathcal{A}$ also share a column in the reference alignment, then we use $CS(\mathcal{A})$ as the average value of 
$C_c$ over the columns of motifs in the reference alignment.

All the experiments with the heuristic of Section~\ref{sec-newheur} were carried out with $M=1,\ldots,4$, $L=20$, 
and $N=200$. The set covers we used were the $\mathcal{I}$ and $\mathcal{S}$ set covers of Table~\ref{tab-1}. Gap 
costs were as given in Table~\ref{tab-2} for each of \texttt{BLOSUM62} \cite{hh92}, \texttt{PAM250} \cite{dso78}, and 
\texttt{VTML160} \cite{msv02} as the substitution matrix.

\begin{table}
\begin{center}
\caption{Gap initialization and extension costs. Sources: \cite{vea95} for \texttt{BLOSUM62} and \texttt{PAM250}; \cite{gb02} for \texttt{VTML160}.}
\begin{tabular}{lrrrr}
\hline
& \multicolumn{4}{c}{Gap costs} \\
\cline{2-5} 
Substitution matrix & \multicolumn{2}{c}{Global alignments} &  \multicolumn{2}{c}{Local alignments} \\
\cline{2-5}
& Initialization & Extension & Initialization & Extension \\
\hline
\texttt{BLOSUM62} & $7.5$ & $0.9$ & $8.0$ & $0.5$ \\
\texttt{PAM250} & $11.0$ & $0.5$ & $6.0$ & $1.3$ \\
\texttt{VTML160} & $14.0$ & $2.0$ & $14.0$ & $2.0$ \\
\hline
\end{tabular}
\label{tab-2}
\end{center}
\end{table}

\begin{figure}
\begin{center}
\begin{tabular}{c}
\includegraphics[scale=0.83]{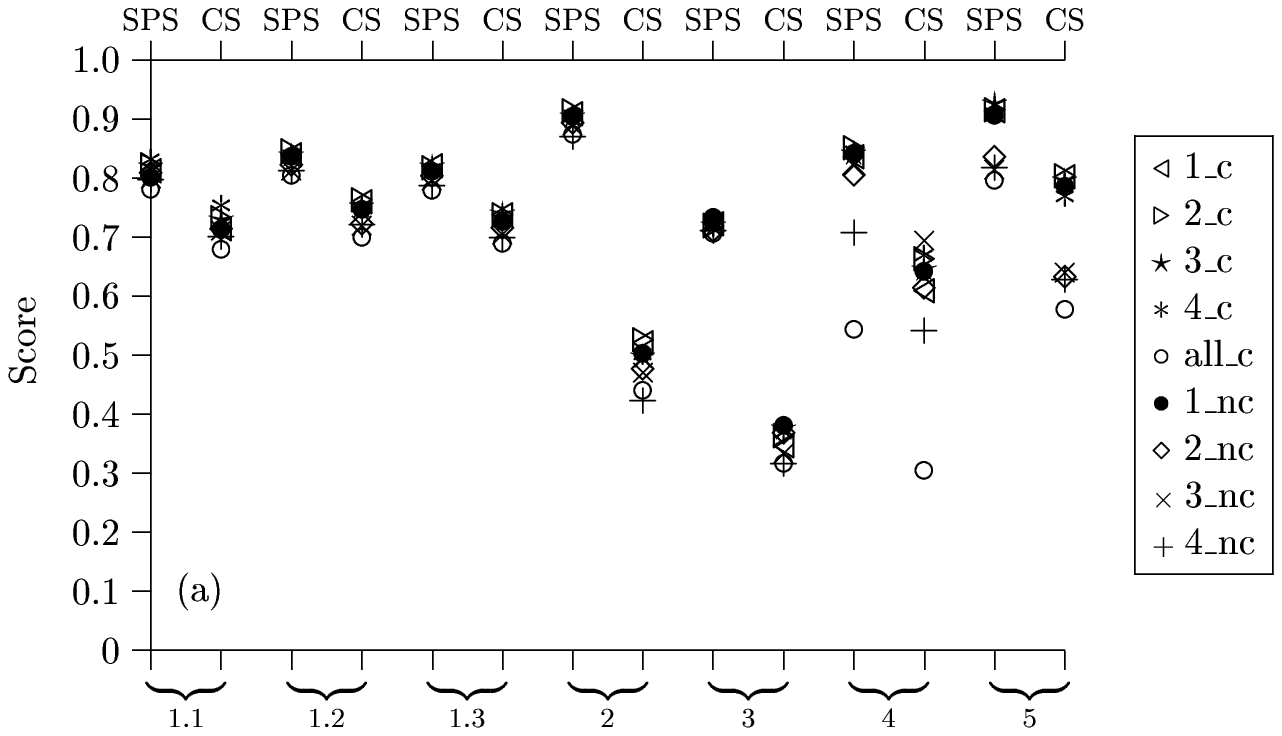} \\
\\ 
\includegraphics[scale=0.83]{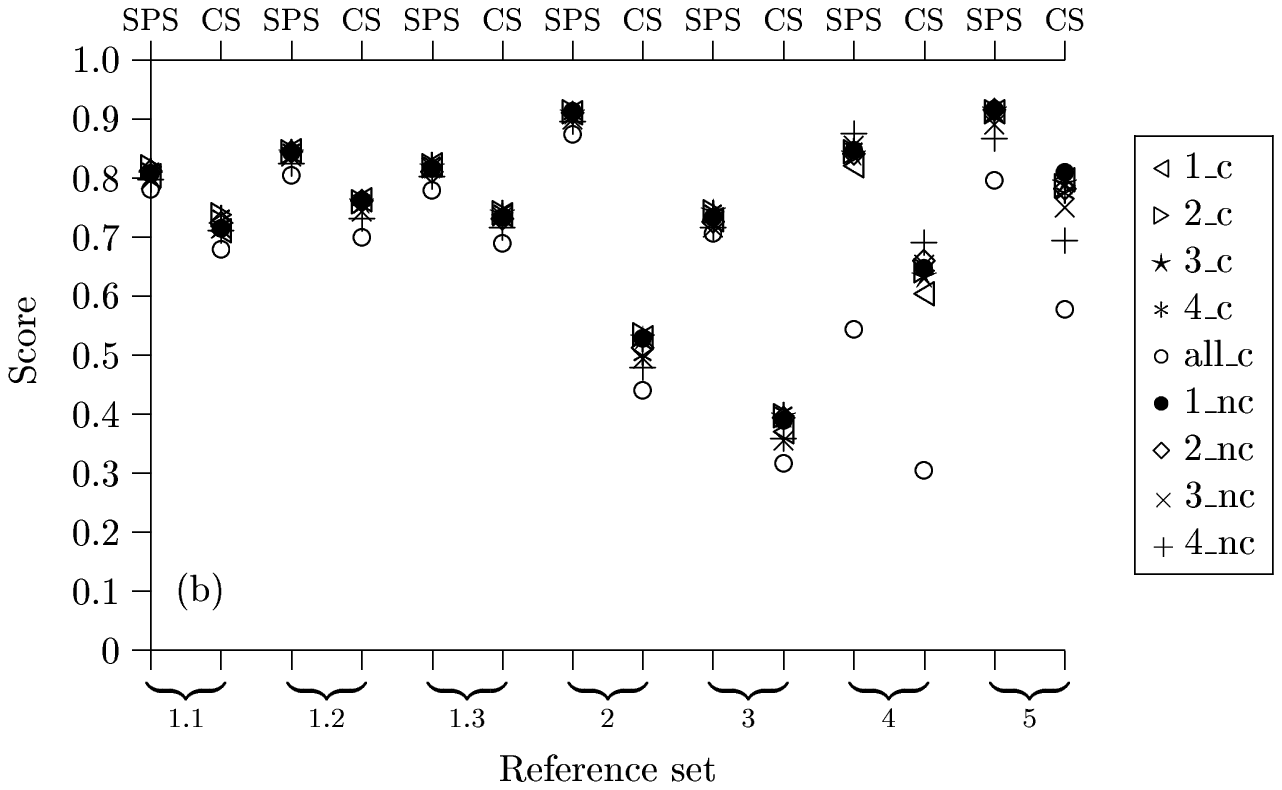} \\
\end{tabular}
\caption{Average scores for the heuristic of Section~\ref{sec-newheur} with the \texttt{BLOSUM62} substitution matrix and the 
$\mathcal{I}$ (a) or $\mathcal{S}$ (b) set cover.}
\label{fig-3}
\end{center}
\end{figure}

\begin{figure}
\begin{center}
\begin{tabular}{c}
\includegraphics[scale=0.83]{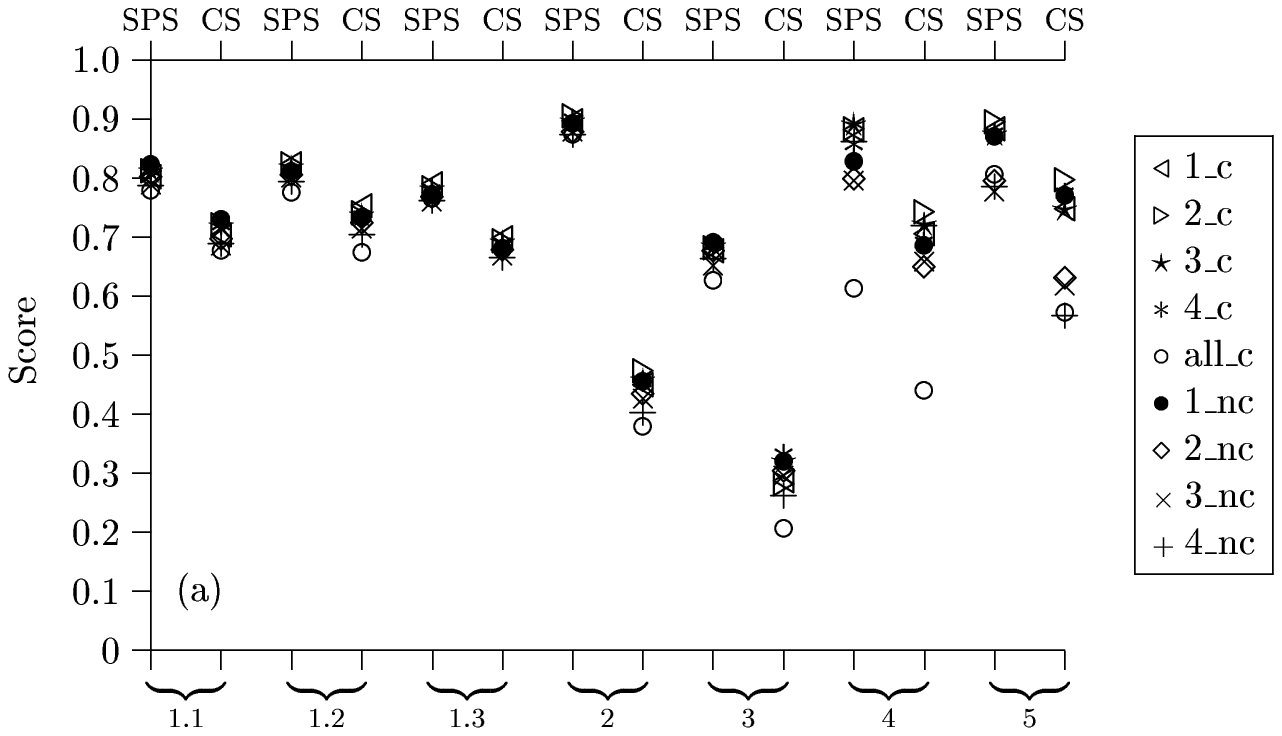} \\
\\
\includegraphics[scale=0.83]{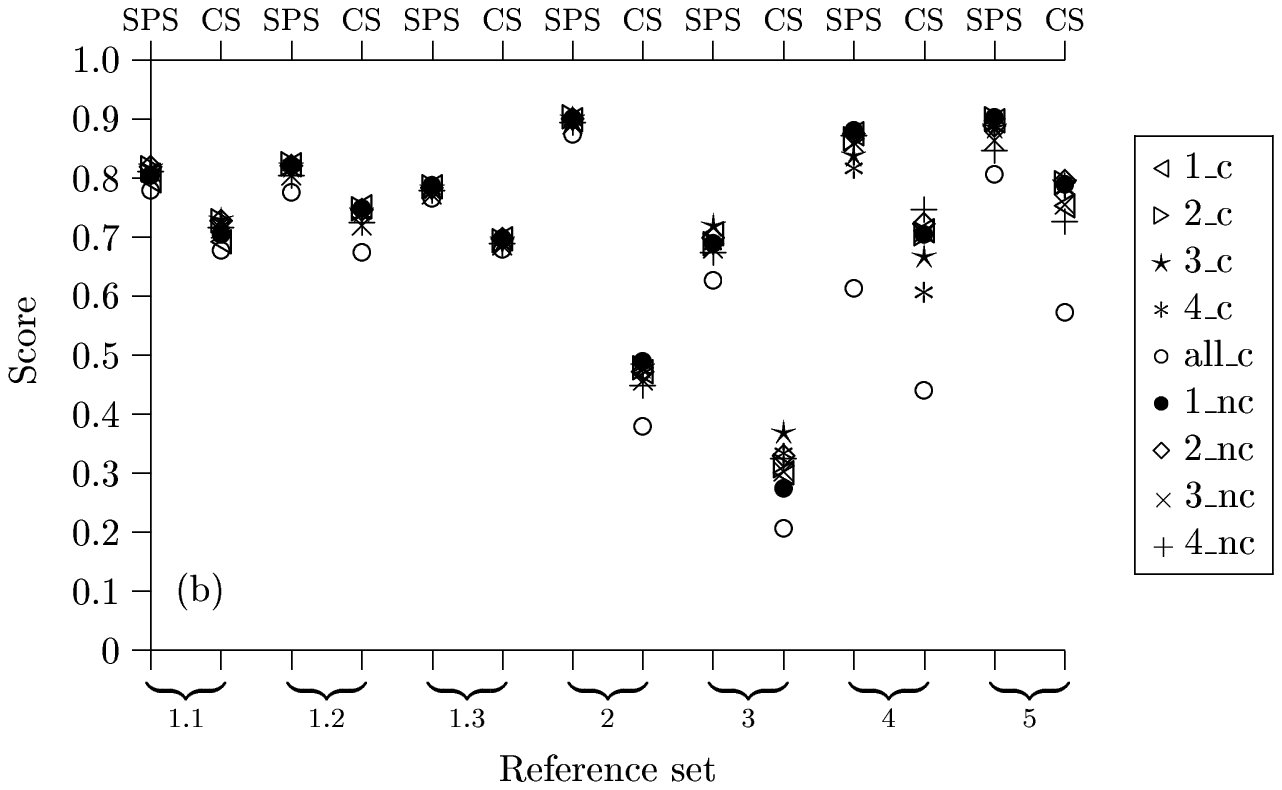} \\
\end{tabular}
\caption{Average scores for the heuristic of Section~\ref{sec-newheur} with the \texttt{PAM250} substitution matrix and the 
$\mathcal{I}$ (a) or $\mathcal{S}$ (b) set cover.}
\label{fig-4}
\end{center}
\end{figure}

\begin{figure}
\begin{center}
\begin{tabular}{c}
\includegraphics[scale=0.83]{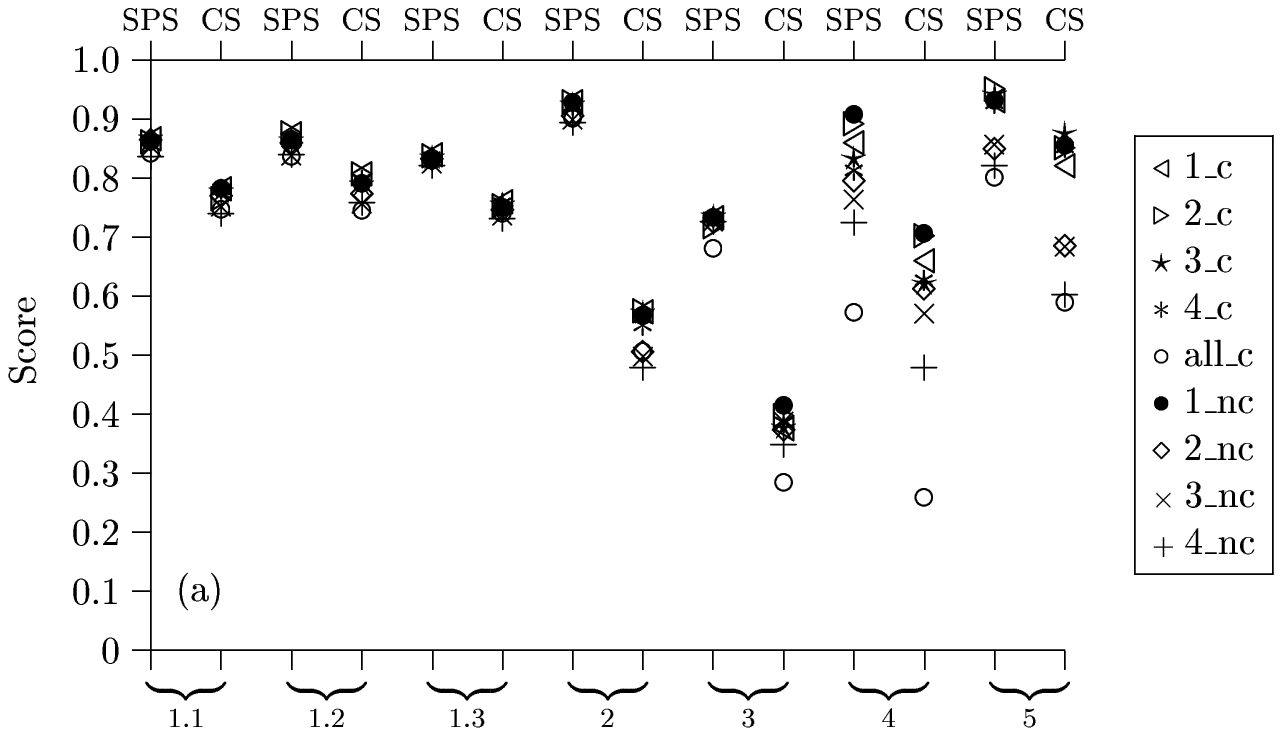} \\
\\
\includegraphics[scale=0.83]{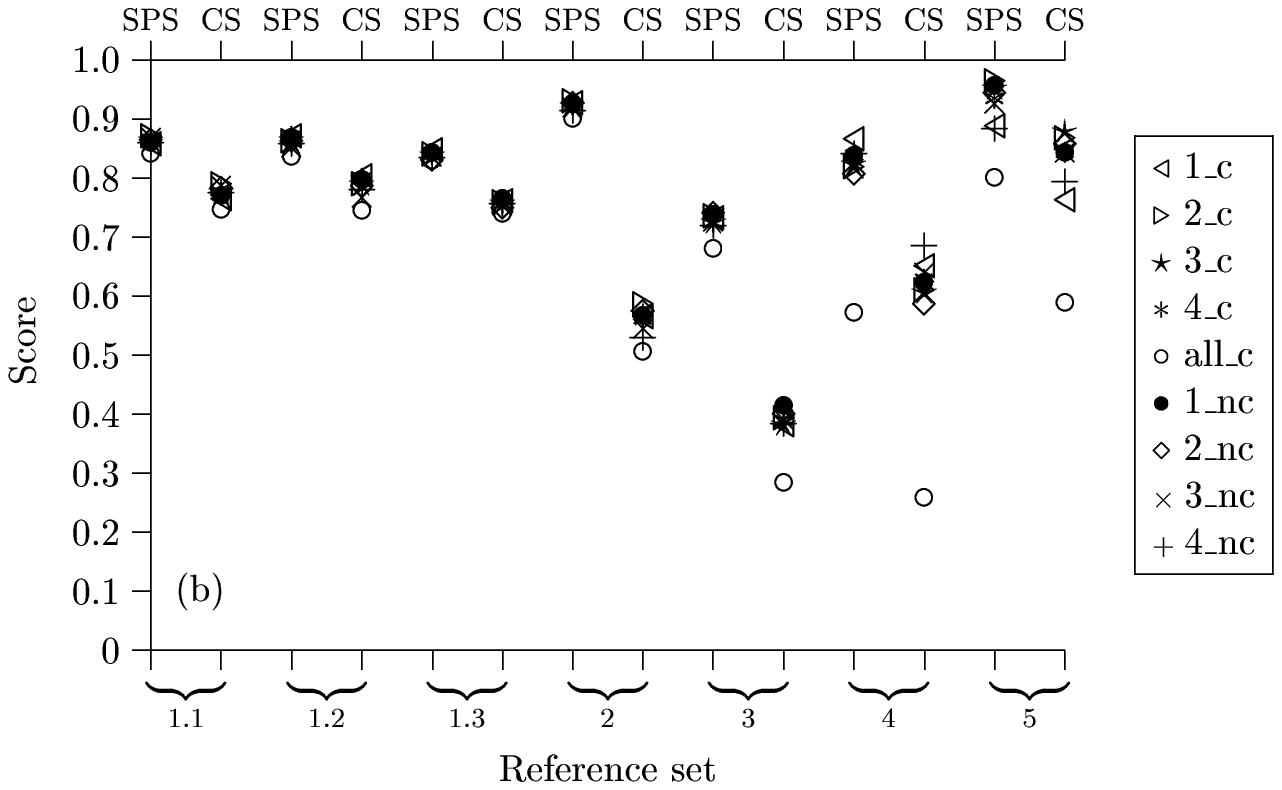} \\
\end{tabular}
\caption{Average scores for the heuristic of Section~\ref{sec-newheur} with the \texttt{VTML160} substitution matrix and the 
$\mathcal{I}$ (a) or $\mathcal{S}$ (b) set cover.}
\label{fig-5}
\end{center}
\end{figure}

We divide our presentation of results into two groups. The first group is given in Figures~\ref{fig-3}--\ref{fig-5}, 
respectively for the \texttt{BLOSUM62}, \texttt{PAM250}, and \texttt{VTML160} matrices, and also in Figure~\ref{fig-6} (see 
below). In each of Figures~\ref{fig-3}--\ref{fig-5}, part (a) refers to the $\mathcal{I}$ set cover, part (b) to the 
$\mathcal{S}$ set cover. Also, figure legends are given as ``x\_y,'' where ``x'' is either the value of $M$ or else the word 
``all,'' in this case indicating that the full suffix-set tree was constructed, and ``y'' is either ``c'' or ``nc'' (for 
``compact'' and ``non-compact''), indicating respectively whether the tree was constructed according to
Steps~\ref{algtree-step-1}--\ref{algtree-step-3} of Section~\ref{sec-newtree} or by the alternative procedure indicated in that 
section (results that would be labeled ``all\_nc'' are altogether absent, since in the case of the $\mathcal{I}$ set cover 
the full tree as constructed by the alternative procedure is large to the extent of rendering the entire approach impractical).

What we show in each figure are average SPS and CS values inside each of the BAliBASE reference sets we considered. 
Clearly, the choices labeled ``2\_c'' emerge from Figures~\ref{fig-3}--\ref{fig-5} as the best alternatives nearly always. 
When we group the results of these choices together for the various combinations of substitution matrix and set cover, 
we obtain what is depicted in Figure~\ref{fig-6}. Clearly, the \texttt{VTML160}-$\mathcal{S}$ pair appears as a first
choice most often, with the only noteworthy exception of Reference Set~4, in which case it seems best to use the pair
\texttt{PAM250}-$\mathcal{I}$.

\begin{figure}
\begin{center}
\includegraphics[scale=0.83]{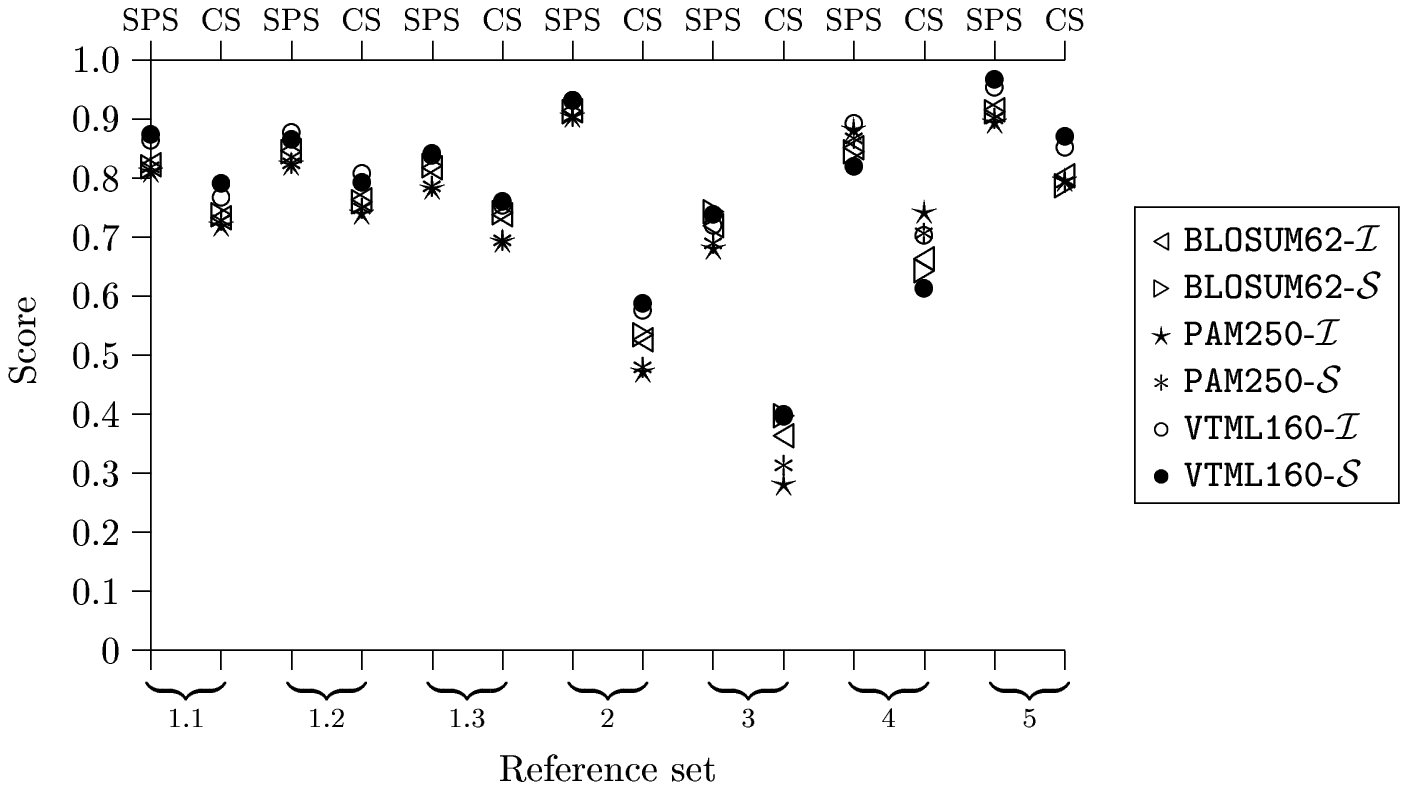} 
\caption{Average scores for the ``2\_c'' cases of Figures~\ref{fig-3}--\ref{fig-5}.}
\label{fig-6}
\end{center}
\end{figure}

Still with regard to Figures~\ref{fig-3}--\ref{fig-5}, it is worth noting that the cases labeled ``all'' or ``nc'' are 
consistently among the worst possibilities. Referring back to our discussion in Section~\ref{sec-newtree}, we see that this 
trend favors the use of the techniques that produce the smaller suffix-set trees (viz.\ prefix-length bounding and
Steps~\ref{algtree-step-1}--\ref{algtree-step-3} as presented in that section). What this means is that, in addition to the 
generally superior scores, we are also to expect a more efficient use of computational resources as the suffix-set tree is constructed. 

Our second group of results refers to comparing the heuristic of Section~\ref{sec-newheur} to the five competing approaches 
mentioned earlier. The approaches that predate the introduction of the BAliBASE suite were run with default parameters (this is the 
case of CLUSTAL W, DIALIGN, and PRRN), while the others, having appeared with experimental results on the BAliBASE suite when first 
published, were run with their best parameter choices (this applies to T-COFFEE and to MAFFT). As for our heuristic, the 
results we use in the comparison are those singled out above as the champions of Figure~\ref{fig-6}.

\begin{figure}
\begin{center}
\includegraphics[scale=0.83]{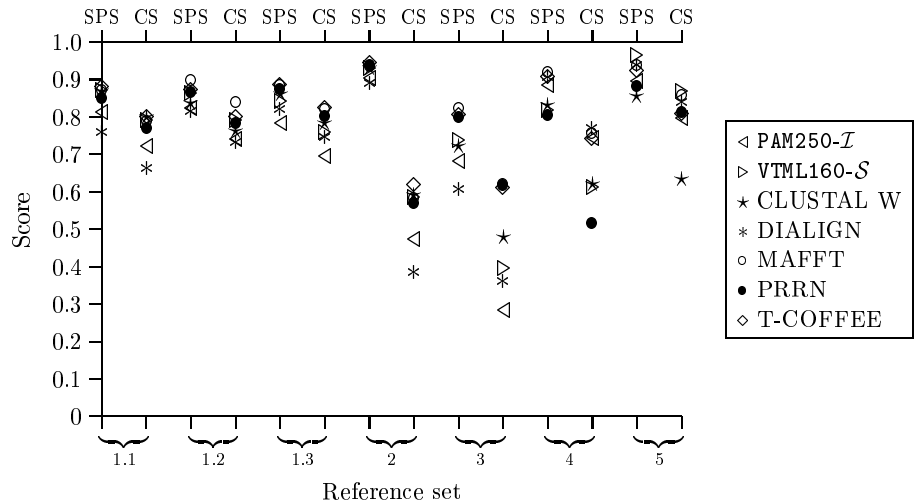} \\
\caption{Average scores for the two best cases of Figure~\ref{fig-6} and the five competing approaches.}
\label{fig-7}
\end{center}
\end{figure}

Comparative results are given in Figure~\ref{fig-7}, where the same style of Figures~\ref{fig-3}--\ref{fig-6} is used, 
and in Table~\ref{tab-3}, where total running times are given.\footnote{All running times were obtained on an Intel Pentium 4
processor running at $1.8$ GHz with $1$ Gbytes of main memory.} It is clear from Figure~\ref{fig-7} that no absolute best can 
be identified throughout all the reference sets. As we examine the reference sets individually, though, we see that at least 
one of the two substitution-matrix, set-cover pairs used with our heuristic is in general competitive with the best contender. Noteworthy 
situations are the superior performance of our heuristic on Reference Set~5, and also its weak performance on Reference Set~3.
As for the running times given in Table~\ref{tab-3}, our current implementation is seen to perform competitively as well 
when compared to the others.

\begin{table}
\begin{center}
\caption{Running times for the two best cases of Figure~\ref{fig-6} and the five competing approaches.}
\begin{tabular}{lr}
\hline
Approach & Time (seconds) \\
\hline
\texttt{PAM250}-$\mathcal{I}$ & 2058.76 \\
\texttt{VTML160}-$\mathcal{S}$ & 1843.27 \\
CLUSTAL W & 111.30 \\
DIALIGN & 338.05 \\
MAFFT & 95.59 \\
PRRN & 2269.84 \\
T-COFFEE & 4922.50 \\
\hline
\end{tabular}
\label{tab-3}
\end{center}
\end{table}

\section{Concluding remarks} \label{sec-concrem}

We have in this paper introduced a new heuristic for the multiple alignment of a set of sequences. The new heuristic is 
based on a set cover of the residue alphabet of the sequences, and also on finding significant subsequence blocks to be combined 
and eventually yield the desired alignment. Central to our heuristic is the notion of a suffix-set tree, a new 
data structure that generalizes the well-known suffix tree of a set of sequences. Even though this tree can conceivably 
have a number of nodes that is exponential in the largest of the sequence lengths, we have demonstrated by means of 
experiments on the BAliBASE suite that best results are achieved when the tree is only partially constructed, in which 
case its number of nodes becomes polynomial in the size of the set cover.

Our experiments on the BAliBASE suite indicate that the new heuristic is competitive when compared to a selection of 
existing strategies. This is true both in terms of motif-constrained versions of the BAliBASE standard scores and also in terms of running times for 
completion. As we focus in a more detailed fashion on individual performance results, our heuristic occasionally 
outperforms all others, and equally occasionally falls behind them. We find it, then, to be a relevant approach, one that 
sheds new light on the usefulness of residue-alphabet set covers.

Many of our heuristic's details can still undergo improvements that go beyond the mere search for a more efficient 
implementation. One possibility is clearly the use of potentially better pairwise alignments, both global and local, when they are 
needed as described in Section~\ref{sec-newheur}. This possibility is already exploited by T-COFFEE, which not only 
employs a position-specific score matrix, but also (as in the implementation used for our comparisons) uses CLUSTAL W to 
obtain global pairwise alignments and LALIGN \cite{hm91} to obtain a set of highly significant local pairwise alignments. 
We also see improvement possibilities in the block- and alignment-extension methods described at the ends of Sections~\ref{sec-newheur-blocks}
and \ref{sec-newheur-alignment}, respectively. In these two occasions, sequences or subsequences are considered in nonincreasing
order of their lengths, which of course is an approach simple to the point of in no way taking into account the biological
significance of the sequences or subsequences.

It is also apparent from our presentation of the heuristic in Section~\ref{sec-newheur} that several options exist for 
many of its building parts. This refers not only to choosing parameter values but also to selecting auxiliary algorithms 
at several points. Whether better choices exist in terms of yielding even more significant alignments, and doing 
it perhaps faster as well, remains to be verified.

\subsection*{Acknowledgments}

The authors acknowledge partial support from CNPq, CAPES, and a FAPERJ BBP grant. They also thank NACAD/UFRJ, on whose
high-performance systems the data for Figure~\ref{fig-2} were obtained.

\bibliography{paper}

\begin{thebibliography}{10}

\bibitem{bttp01}
A.~Bahr, J.~D. Thompson, J.-C. Thierry, and O.~Poch.
\newblock {BA}li{BASE} ({B}enchmark {A}lignment data{BASE}): enhancements for
  repeats, transmembrane sequences and circular permutations.
\newblock {\em Nucleic Acids Research}, 29:323--326, 2001.

\bibitem{bb01}
P.~Baldi and S.~Brunak.
\newblock {\em Bioinformatics}.
\newblock The MIT Press, Cambridge, MA, 2001.

\bibitem{bv01}
P.~Bonizzoni and G.~Della Vedora.
\newblock The complexity of multiple sequence alignment with {SP}-score that is
  a metric.
\newblock {\em Theoretical Computer Science}, 259:63--79, 2001.

\bibitem{bjeg98}
A.~Brazma, I.~Jonassen, I.~Eidhammer, and D.~Gilbert.
\newblock Approaches to the automatic discovery of patterns in biosequences.
\newblock {\em Journal of Computational Biology}, 5:279--305, 1998.

\bibitem{ctrv02}
N.~Cannata, S.~Toppo, C.~Romualdi, and G.~Valle.
\newblock Simplifying amino acid alphabets by means of a branch and bound
  algorithm and substitution matrices.
\newblock {\em Bioinformatics}, 18:1102--1108, 2002.

\bibitem{clrs01}
T.~H. Cormen, C.~E. Leiserson, R.~L. Rivest, and C.~Stein.
\newblock {\em Introduction to Algorithms}.
\newblock The MIT Press, Cambridge, MA, second edition, 2001.

\bibitem{dso78}
M.~O. Dayhoff, R.~M. Schwartz, and B.~C. Orcutt.
\newblock A model of evolutionary change in proteins.
\newblock In M.~O. Dayhoff, editor, {\em Atlas of Protein Sequence and
  Structure}, volume 5, supplement\ 3, pages 345--352. National Biomedical
  Research Foundation, Washington, DC, 1978.

\bibitem{d81}
R.~F. Doolittle.
\newblock Similar amino acid sequences: chance or common ancestry?
\newblock {\em Science}, 214:149--159, 1981.

\bibitem{dms98}
A.~Dress, B.~Morgenstern, and J.~Stoye.
\newblock The number of standard and of effective multiple alignments.
\newblock {\em Applied Mathematics Letters}, 11:43--49, 1998.

\bibitem{dekm98}
R.~Durbin, S.~Eddy, A.~Krogh, and G.~Mitchison.
\newblock {\em Biological Sequence Analysis}.
\newblock Cambridge University Press, Cambridge, UK, 1998.

\bibitem{ejt04}
I.~Eidhammer, I.~Jonassen, and W.~R. Taylor.
\newblock {\em Protein Bioinformatics}.
\newblock John Wiley \& Sons, Chichester, UK, 2004.

\bibitem{gs97}
R.~Giegerich and S.~Kurtz.
\newblock From {U}kkonen to {M}c{C}reight and {W}einer: a unifying view of
  linear-time suffix tree construction.
\newblock {\em Algorithmica}, 19:331--353, 1997.

\bibitem{g96}
O.~Gotoh.
\newblock Significant improvement in accuracy of multiple protein sequence
  alignments by iterative refinement as assessed by reference to structural
  alignments.
\newblock {\em Journal of Molecular Biology}, 264:823--838, 1996.

\bibitem{g99}
O.~Gotoh.
\newblock Multiple sequence alignment: algorithms and applications.
\newblock {\em Advances in Biophysics}, 36:159--206, 1999.

\bibitem{gb02}
R.~E. Green and S.~E. Brenner.
\newblock Bootstrapping and normalization for enhanced evaluations of pairwise
  sequence comparison.
\newblock {\em Proceedings of the IEEE}, 90:1834--1847, 2002.

\bibitem{g97}
D.~Gusfield.
\newblock {\em Algorithms on Strings, Trees, and Sequences}.
\newblock Cambridge University Press, Cambridge, UK, 1997.

\bibitem{hh92}
S.~Henikoff and J.~G. Henikoff.
\newblock Amino acid substitution matrices from protein blocks.
\newblock {\em Proceedings of the National Academy of Sciences USA},
  89:10915--10919, 1992.

\bibitem{hm91}
X.~Huang and W.~Miller.
\newblock A time-efficient, linear-space local similarity algorithm.
\newblock {\em Advances in Applied Mathematics}, 12:337--357, 1991.

\bibitem{i97}
T.~R. Ioerger.
\newblock The context-dependence of amino acid properties.
\newblock In {\em Proceedings of the Fifth International Conference on
  Intelligent Systems for Molecular Biology}, pages 157--166, 1997.

\bibitem{jz81}
M.~A. Jim{\'{e}}nez-Monta{\~{n}}o and L.~Zamora-Cortina.
\newblock Evolutionary model for the generation of amino acid sequences and its
  application to the study of fragments of mammal-hemoglobin chains.
\newblock In {\em Proceedings of the Seventh International Biophysics
  Congress}, 1981.

\bibitem{j01}
W.~Just.
\newblock Computational complexity of multiple sequence alignment with
  {SP}-score.
\newblock {\em Journal of Computational Biology}, 8:615--623, 2001.

\bibitem{kh01}
K.~Karplus and B.~Hu.
\newblock Evaluation of protein multiple alignments by {SAM}-{T}99 using the
  {BA}li{BASE} multiple alignment test set.
\newblock {\em Bioinformatics}, 17:713--720, 2001.

\bibitem{kmkm02}
K.~Katoh, K.~Misawa, K.~Kuma, and T.~Miyata.
\newblock {MAFFT}: a novel method for rapid multiple sequence alignment based
  on fast {F}ourier transform.
\newblock {\em Nucleic Acids Research}, 30:3059--3066, 2002.

\bibitem{k96b}
T.~M. Klingler.
\newblock {\em Structural Inference from Correlations in Biological Sequences}.
\newblock PhD thesis, Program in Medical Informatics, Stanford University,
  1996.

\bibitem{ls02}
T.~Lassmann and E.~L.~L. Sonnhammer.
\newblock Quality assessment of multiple alignment programs.
\newblock {\em FEBS Letters}, 529:126--130, 2002.

\bibitem{ltptp01}
O.~Lecompte, J.~D. Thompson, F.~Plewniak, J.-C. Thierry, and O.~Poch.
\newblock Multiple alignment of complete sequences ({MACS}) in the post-genomic
  era.
\newblock {\em Gene}, 270:17--30, 2001.

\bibitem{lfww03}
T.~P. Li, K.~Fan, J.~Wang, and W.~Wang.
\newblock Reduction of protein sequence complexity by residue grouping.
\newblock {\em Protein Engineering}, 16:323--330, 2003.

\bibitem{lb93}
C.~D. Livingstone and G.~J. Barton.
\newblock Protein sequence alignments: a strategy for the hierarchical analysis
  of residue conservation.
\newblock {\em Computer Applications in the Biosciences}, 9:745--756, 1993.

\bibitem{m03}
B.~Manthey.
\newblock Non-approximability of weighted multiple sequence alignment.
\newblock {\em Theoretical Computer Science}, 296:179--192, 2003.

\bibitem{m76}
E.~M. McCreight.
\newblock A space-economical suffix tree construction algorithm.
\newblock {\em Journal of the ACM}, 23:262--272, 1976.

\bibitem{mbrzh94}
W.~Miller, M.~Boguski, B.~Raghavachari, Z.~Zhang, and R.~C. Hardison.
\newblock Constructing aligned sequence blocks.
\newblock {\em Journal of Computational Biology}, 1:51--64, 1994.

\bibitem{m95}
G.~Mocz.
\newblock Fuzzy cluster analysis of simple physicochemical properties of amino
  acids for recognizing secondary structure in proteins.
\newblock {\em Protein Science}, 4:1178--1187, 1995.

\bibitem{m99}
B.~Morgenstern.
\newblock {DIALIGN} 2: improvement of the segment-to-segment approach to
  multiple sequence alignment.
\newblock {\em Bioinformatics}, 15:211--218, 1999.

\bibitem{mfdw98}
B.~Morgenstern, K.~Frech, A.~Dress, and T.~Werner.
\newblock {DIALIGN}: finding local similarities by multiple sequence alignment.
\newblock {\em Bioinformatics}, 14:290--294, 1998.

\bibitem{msv02}
T.~M{\"{u}}ller, R.~Spang, and M.~Vingron.
\newblock Estimating amino acid substitution models: a comparison of
  {D}ayhoff's estimator, the resolvent approach and a maximum likelihood
  method.
\newblock {\em Molecular Biology and Evolution}, 19:8--13, 2002.

\bibitem{nfjwn96}
D.~Naor, D.~Fischer, R.~L. Jernigan, H.~J. Wolfson, and R.~Nussinov.
\newblock Amino acid pair interchanges at spatially conserved locations.
\newblock {\em Journal of Molecular Biology}, 256:924--938, 1996.

\bibitem{nw70}
S.~B. Needleman and C.~D. Wunsch.
\newblock A general method applicable to the search for similarities in the
  amino acid sequence of two proteins.
\newblock {\em Journal of Molecular Biology}, 48:443--453, 1970.

\bibitem{nrd02}
H.~B. {Nicholas Jr}., A.~J. Ropelewski, and D.~E. {Deerfield II}.
\newblock Strategies for multiple sequence alignment.
\newblock {\em Biotechniques}, 32:572--591, 2002.

\bibitem{n02}
C.~Notredame.
\newblock Recent progress in multiple sequence alignment: a survey.
\newblock {\em Pharmacogenomics}, 3:131--144, 2002.

\bibitem{nhh00}
C.~Notredame, D.~G. Higgins, and J.~Heringa.
\newblock {T}-{C}offee: a novel method for fast and accurate multiple sequence
  alignment.
\newblock {\em Journal of Molecular Biology}, 302:205--217, 2000.

\bibitem{pfr99}
L.~Parida, A.~Floratos, and I.~Rigoutsos.
\newblock An approximation algorithm for alignment of multiple sequences using
  motif discovery.
\newblock {\em Journal of Combinatorial Optimization}, 3:247--275, 1999.

\bibitem{p00}
P.~A. Pevzner.
\newblock {\em Computational Molecular Biology}.
\newblock The MIT Press, Cambridge, MA, 2000.

\bibitem{rfpgp00}
I.~Rigoutsos, A.~Floratos, L.~Parida, Y.~Gao, and D.~Pratt.
\newblock The emergence of pattern discovery techniques in computational
  biology.
\newblock {\em Metabolic Engineering}, 2:159--177, 2000.

\bibitem{sw03}
M.-F. Sagot and Y.~Wakabayashi.
\newblock Pattern inference under many guises.
\newblock In B.~A. Reed and C.~L. Sales, editors, {\em Recent Advances in
  Algorithms and Combinatorics}, pages 245--287. Springer-Verlag, New York, NY,
  2003.

\bibitem{s75}
D.~Sankoff.
\newblock Minimum mutation trees of sequences.
\newblock {\em SIAM Journal on Applied Mathematics}, 28:35--42, 1975.

\bibitem{sm97}
J.~Setubal and J.~Meidanis.
\newblock {\em Introduction to Computational Molecular Biology}.
\newblock PWS Publishing Company, Boston, MA, 1997.

\bibitem{s98}
J.~B. Slowinski.
\newblock The number of multiple alignments.
\newblock {\em Molecular Phylogenetics and Evolution}, 10:264--266, 1998.

\bibitem{ss90}
R.~F. Smith and T.~F. Smith.
\newblock Automatic generation of primary sequence patterns from sets of
  related protein sequences.
\newblock {\em Proceedings of the National Academy of Sciences USA},
  87:118--122, 1990.

\bibitem{sw81}
T.~F. Smith and M.~S. Waterman.
\newblock Identification of commom molecular subsequences.
\newblock {\em Journal of Molecular Biology}, 147:195--197, 1981.

\bibitem{s66}
P.~H. Sneath.
\newblock Relations between chemical structure and biological activity in
  peptides.
\newblock {\em Journal of Theoretical Biology}, 12:157--195, 1966.

\bibitem{sm86}
E.~Sobel and H.~M. Martinez.
\newblock A multiple sequence alignment program.
\newblock {\em Nucleic Acids Research}, 14:363--374, 1986.

\bibitem{s96}
L.~E. Stanfel.
\newblock A new approach to clustering the amino acids.
\newblock {\em Journal of Theoretical Biology}, 183:195--205, 1996.

\bibitem{t86}
W.~R. Taylor.
\newblock The classification of amino acid conservation.
\newblock {\em Journal of Theoretical Biology}, 119:205--218, 1986.

\bibitem{t99}
W.~R. Taylor.
\newblock The properties of amino acids in sequences.
\newblock In M.~J. Bishop, editor, {\em Genetic Databases}, pages 81--103.
  Academic Press, London, UK, 1999.

\bibitem{thg94}
J.~D. Thompson, D.~G. Higgins, and T.~J. Gibson.
\newblock {CLUSTAL} {W}: improving the sensitivity of progressive multiple
  sequence alignment through sequence weighting, position-specific gap
  penalties, and weight matrix choice.
\newblock {\em Nucleic Acids Research}, 22:4673--4680, 1994.

\bibitem{tpp99a}
J.~D. Thompson, F.~Plewniak, and O.~Poch.
\newblock {BA}li{BASE}: a benchmark alignment database for the evaluation of
  multiple alignment programs.
\newblock {\em Bioinfomatics}, 15:87--88, 1999.

\bibitem{tpp99b}
J.~D. Thompson, F.~Plewniak, and O.~Poch.
\newblock A comprehensive comparison of multiple sequence alignment programs.
\newblock {\em Nucleic Acids Research}, 27:2682--2690, 1999.

\bibitem{u95}
E.~Ukkonen.
\newblock On-line construction of suffix trees.
\newblock {\em Algorithmica}, 14:249--260, 1995.

\bibitem{vms99}
A.~Vanet, L.~Marsan, and M.-F. Sagot.
\newblock Promoter sequences and algorithmical methods for identifying them.
\newblock {\em Research in Microbiology}, 150:779--799, 1999.

\bibitem{vvp01}
D.~Voet, J.~G. Voet, and C.~W. Pratt.
\newblock {\em Fundamentals of Biochemistry}.
\newblock John Wiley \& Sons, New York, NY, 2001.

\bibitem{vea95}
G.~Vogt, T.~Etzold, and P.~Argos.
\newblock An assessment of amino acid exchange matrices in aligning protein
  sequences: the twilight zone revisited.
\newblock {\em Journal of Molecular Biology}, 249:816--831, 1995.

\bibitem{wj94}
L.~Wang and T.~Jiang.
\newblock On the complexity of multiple sequence alignment.
\newblock {\em Journal of Computational Biology}, 1:337--348, 1994.

\bibitem{wj90}
M.~S. Waterman and R.~Jones.
\newblock Consensus methods for {DNA} and protein sequence alignment.
\newblock In {\em Methods in Enzymology}, volume 183, pages 221--237. Academic
  Press, 1990.

\bibitem{wsb76}
M.~S. Waterman, T.~F. Smith, and W.~A. Beyer.
\newblock Some biological sequence metrics.
\newblock {\em Advances in Mathematics}, 20:367--387, 1976.

\bibitem{w73}
P.~Weiner.
\newblock Linear pattern matching algorithms.
\newblock In {\em Proceedings of the Fourteenth Symposium on Switching and
  Automata Theory}, pages 1--11, 1973.

\bibitem{wb96}
T.~D. Wu and D.~L. Brutlag.
\newblock Discovering empirically conserved amino acid substitution groups in
  databases of protein families.
\newblock In {\em Proceedings of the Fourth International Conference on
  Intelligent Systems for Molecular Biology}, pages 230--240, 1996.

\bibitem{zhm96}
Z.~Zhang, B.~He, and W.~Miller.
\newblock Local multiple alignment via subgraph enumeration.
\newblock {\em Discrete Applied Mathematics}, 71:337--365, 1996.

\bibitem{zj01}
P.~Zhao and T.~Jiang.
\newblock A heuristic algorithm for multiple sequence alignment based on
  blocks.
\newblock {\em Journal of Combinatorial Optimization}, 5:95--115, 2001.

\end{thebibliography}
\bibliographystyle{plain}

\end{document}